# Magnetic moment preservation and emergent Kondo resonance of Co-phthalocyanine on semimetallic Sb(111)

Limin She[1,*], Zhitao Shen[1,*], Zhenyang Xie[1], Limei Wang[1], Yeheng Song[1], Xue-sen Wang[2,1], Yu Jia[1,3,4], Zhenyu Zhang[5,†], Weifeng Zhang[1,‡]

[1]*Key Laboratory for Quantum Matters, and Key Laboratory of Photovoltaic Materials, Henan University, Kaifeng 475004, China*

[2]*Department of Physics, National University of Singapore, 117542, Singapore*

[3]*International Laboratory for Quantum Functional Materials of Henan, Zhengzhou University, Zhengzhou 450003, China*

[4]*Key Laboratory for Special Functional Materials of Ministry of Education, Henan University, Kaifeng 475004, China*

[5]*International Center for Quantum Design of Functional Materials (ICQD), Hefei National Laboratory for Physical Sciences at the Microscale, University of Science and Technology of China, Hefei 230026, China*

Magnetic molecules on surfaces have been widely investigated to reveal delicate interfacial couplings and for potential technological applications. In these endeavors, one prevailing challenge is how to preserve or recover the molecular spins, especially on highly metallic substrates that can readily quench the magnetic moments of the ad-molecules. Here we use scanning tunneling microscopy/spectroscopy to exploit the semimetallic nature of antimony and observe, surprisingly yet pleasantly, that the spin of Co-phthalocyanine is well preserved on Sb(111), as unambiguously evidenced by the emergent strong Kondo resonance across the molecule. Our first-principles calculations further confirm that the optimal density of states near the Fermi level of the semimetal is a decisive factor, weakening the overall interfacial coupling, while still ensuring sufficiently effective electron-spin scattering in the many-body system. Beyond isolated ad-molecules, we discover that each of the magnetic moments in a molecular dimer or a densely packed island is distinctly preserved as well, rendering such molecular magnets immense potentials for ultra-high density memory devices.

Molecular magnets are appealing elemental building blocks for realizing nanoscale information storage and spintronic devices [1-6]. When adsorbed, their electronic and magnetic properties can be substantially modified by the hosting substrates, with the underlying nature of interfacial couplings sensitively dependent on the specific molecule/substrate combinations [7-21]. For a given system, a comprehensive understanding of the fundamental interfacial processes is a prerequisite in gaining precise control of the magnetic properties of the ad-molecule for desirable functionalities.

In the endeavor of exploiting ad-molecule based spintronic devices, a widely studied species is Co-phthalocyanine (CoPc), with spin 1/2. However, the intrinsic magnetic moment of CoPc is quenched when adsorbed on highly metallic substrates such as Cu(111), Ag(111), Au(111), and even on ferromagnetic Fe(110), and the underlying reason is mainly attributed to pronounced net charge transfer or strong interfacial bonding [7,17,18,22-24]. To date, various ingenious approaches, including bond-selective dehydrogenation and insertion of a buffer layer, have been devised to recover the magnetic moment of the adsorbed CoPc [7,18,25]. Theoretical studies have also revealed that the interfacial hybridization strengths can strongly affect the magnetic properties of CoPc, and can be substantially tuned by changing the molecule-substrate separation for a given system [17,26]. Beyond such prevailing interface modification schemes, it is even more attractive to identify substrates that can naturally preserve the intrinsic moment of CoPc while also embody rich many-body physics of electron-spin interfacial coupling to be elucidated.

In this Letter, we combine experimental and theoretical approaches to establish the vital role of an optimal density of states around the Fermi level of the substrate in preserving the intrinsic spin of the adsorbed CoPc molecule. Using scanning tunneling microscopy/spectroscopy (STM/STS), we find, surprisingly and yet pleasantly, a CoPc molecule on the semimetallic Sb(111) substrate exhibits spatially extended Kondo resonance even at the relatively high temperature of ~220 K, an unambiguous indication that its molecular spin is well preserved upon adsorption. Our first-principles calculations within density functional theory (DFT) further show that the interfacial hybridization is optimized, allowing the unpaired $d_{z^2}$ orbital of CoPc to endure the adsorption and interfacial charge transfer while also ensure sufficiently effective spin-flip scattering of the many-body electrons of the semimetallic Sb(111). Moreover, the DFT calculations reveal that the interfacial charge transfer induces intramolecular reorganization of the spin density. Beyond isolated molecules, our STS measurements further resolve that each of the CoPc molecules in a closely coupled molecular dimer or a densely packed molecular island distinctly exhibits its own Kondo resonance in the presence of antiferromagnetic intermolecular RKKY coupling, laying the foundation for the CoPc assembly to be utilized in ultrahigh-density spin devices.

The experiments were carried out by using a commercial STM (Unisoku 1300, Japan) with a base pressure better than $1.0 \times 10^{-10}$ Torr [see Section I in SM [27]]. Figure 1(a) shows a high resolution STM image of three isolated CoPc molecules adsorbed at the bridge sites of Sb(111), exhibiting nearly flat-lying adsorption configurations. At a refined level, there exist two orientations of the ad-molecules, $S_1$, where the molecular lobes point to the $[\bar{2}11]$ and $[01\bar{1}]$ directions, and $S_2$, rotated by ~30° from $S_1$. These configurations are consistent with previous identifications [28]. To understand their electronic properties, differential conductance spectra (*dI/dV*) were taken at the Co ions of both configurations at 5 K. As shown in Fig. 1(b), the narrow peaks in the *dI/dV* spectra are located at ~10 and 3 meV below $E_F$ for $S_1$ and $S_2$, respectively. Here we note that, since the peaks associated with the CoPc-$d_{z^2}$ orbital-mediated tunneling are typically located about 0.1 - 0.4 eV below $E_F$ with a width of ~100 meV, the sharp peaks very close to $E_F$ observed here must be due to different physical origin(s) [7,29].

As a natural interpretation, we attribute the above peaks near $E_F$ to the Kondo resonances, originating from the multiple spin-flip scatterings of the conduction electrons at the intrinsic spins of the molecules. Such an interacting many-body system can be well described by the Fano model, and the lineshape of the *dI/dV* spectra in Fig. 1(b) can be fitted by the Fano formula as [30-33],

$$\frac{dI}{dV} \propto \frac{(\varepsilon+q)^2}{1+\varepsilon^2}, \quad \varepsilon = \frac{eV-\varepsilon_0}{\Gamma},$$

where $q$ is the asymmetric Fano factor, $\varepsilon_0$ is the energy shift of the Kondo resonance center from $E_F$, and $\Gamma$ is the half width of the resonance. For the $S_1$ and $S_2$ configurations, we acquire the average values of the physical parameters to be $\Gamma = 28.6 \pm 2$ and $26.1 \pm 2$ meV, respectively. The Kondo temperatures are determined to be $T_K = 234.4 \pm 17$ K ($S_1$) and $T_K = 213.8 \pm 17$ K ($S_2$) according to the relationship $\Gamma = \sqrt{(\pi k_B T)^2 + 2(k_B T_K)^2}$, where $T$ is the substrate temperature, and $k_B$ is the Boltzmann constant [32]. The observed $T_K$ is noticeably higher than that of Co atoms on Sb(111) and dehydrogenated CoPc (*d*-CoPc) on Au(111) [7,34,35]. Here it is worthwhile to note that Co atoms on Sb(111) can still exhibit Kondo resonance because such atoms possess spins of 3/2, more robust against quenching due to charge transfer [36,37].

To reveal the spatial distribution of the Kondo resonance, *dI/dV* maps were acquired at sample voltages corresponding to the peak energy of the resonance, as shown in Figs. 1(c) and 1(d). Compared with the STM topographies [see Figs. S1(c) and S1(d) in SM [27]], we find that the Kondo resonance of CoPc is not only located at the Co ion, but also extends to the molecular pyrrole groups. This spatial distribution of the Kondo resonance is often driven by the *π-d* interaction or the reorganization of the molecular charge density, which is further verified by a series of *dI/dV* spectra acquired along different axes over the CoPc molecule [8,20]. As shown in Figs. 1(e)-1(h), the Kondo resonance decays slowly as the STM tip moves away from the center of the molecule along two different directions. For a given configuration, the half widths of the Kondo resonance are almost identical at Co and pyrrole ring [see Figs. S2(a) and S2(b) in SM [27]], indicating nearly uniform Kondo temperatures as well.

To gain deeper insight into the origin of the resonance peak near $E_F$, we have performed first-principles studies of the CoPc/Sb(111) system [see Section II in SM [27]]. In doing so, we construct the initial modes of CoPc molecules at the bridge sites of Sb(111) based on the STM results. Two stable adsorption configurations are identified, in which the CoPc molecules lie largely flat, but slightly concaved. Figures 2(a)-2(c) demonstrate the calculated spin-polarized partial densities of states (PDOS) on the Co ion and phthalocyanine (Pc) ligand in the free CoPc, $S_1$, and, $S_2$ configurations, respectively, where the majority and minority spins of the $d$ orbitals are clearly unevenly occupied. Especially, all the $d_{z^2}$ orbitals are singly occupied, contributing significantly to the magnetic moment of the CoPc/Sb(111) composite. It is also noted that the $d_{z^2}$ orbital near $E_F$ in either configuration becomes broader when the molecule is adsorbed on Sb(111), indicating strong coupling between the discrete $d_{z^2}$ spin state and surrounding conduction electron continuums of the Fano system.

In order to further unravel the nature of the interfacial coupling, we calculate the PDOS of bare Sb(111) and CoPc/Sb(111) complexes [Figs. 2(d)-2(f)]. We first note that, the DOS of the complex near $E_F$ is significantly lower than that of Cu(111), explaining why in the latter case the magnetic moment of CoPc is readily quenched [26]. For our present system, we propose that the conduction electrons of the Sb(111) around $E_F$ behave like a perfect "glue" in fine tuning the hybridization strength, such that the molecular spin is preserved due to insignificant charge transfer and yet still effectively coupled to the itinerant electrons to ensure multiple spin-flip scatterings. Our calculations indeed show that the magnetic moments are 0.962 ($S_1$) and 0.943 $\mu_B$ ($S_2$), only slightly less than that of the free CoPc (0.998 $\mu_B$). The average distances between the Co ion and the Sb surface are about 3.37 ($S_1$) and 3.31 Å ($S_2$), agreeing well with the STM data [see Figs. S3(c) and S3(d) in SM [27]]. Obviously, the Co-Sb distances are larger than that of CoPc on other metal substrates, especially those highly metallic ones [22,23,26,38,39]. Moreover, our selective test calculations show that the magnetic moment of CoPc is quenched rapidly as the Co-Sb distance decreases [see Fig. S4 in SM [27]]. On the basis of these calculations, we conclude that the distance between Co and substrate as well as the interfacial hybridization strengths are mainly controlled by the DOS of the host substrate around $E_\text{F}$, which is crucial to preserve the molecular spin and drive the CoPc-substrate composites into the Kondo regime.

At a finer level, our calculated spin density maps [Figs. 3(a)-3(d)] of the two configurations show clear spin spreading from the central Co to the pyrrole rings, as also observed in the STM measurements of the Kondo resonance. In addition, the spin of the ligand is antiparallel to the Co spin, resulting in a minor decrease in the molecular magnetic moment. Nevertheless, the Kondo resonance in the CoPc-Sb(111) system is not suppressed by the presence of the apparent intramolecular antiferromagnetic coupling.

Our calculated differential electronic density maps of the CoPc-Sb(111) complexes are shown in Figs. 3(e)-3(h). For a given configuration, there exists obvious charge transfer from the surface to the molecule, resulting in the spin-active pyrrole ring due to charge redistribution of the Co ion and ligand. Through Barder charge analysis, we find that ~0.19 ($S_1$) and 0.22 $e$ ($S_2$) are transferred from the Sb(111) substrate to the CoPc molecules [40]. Moreover, the results also show that the 3$d$ orbital occupations ($S_1$: 7.31 and $S_2$: 7.34) are similar to that of the free CoPc

(7.26), indicating that the molecule has a $3d^7$ Co(II) center with spin 1/2. Again, we attribute the robust spin of CoPc to the weak interfacial hybridization on the semimetallic Sb(111).

Going beyond isolated ad-molecules, Fig. 4(a) contrasts the measured Kondo temperature taken from an isolated molecule, a molecular dimer, a trimer, and a much larger island, showing a clear increase in $T_K$ with the assembly size for the smaller assemblies [see Figs. S5(a) and S5(b) in SM [27]]. Figure 4(b) also shows the *dI/dV* spectra of CoPc in the molecular island measured at 5 and 78 K, yielding $\Gamma = 32.9 \pm 2$ and $39.5 \pm 3$ meV, respectively. The corresponding Kondo temperatures are given as $T_K = 269.8 \pm 17$ and $272.5 \pm 30$ K. Such an obvious temperature dependence further acts as a fingerprint of the Kondo resonance [32]. We also note that the width of the resonance peak of CoPc in the molecular island is slightly broader than that of an isolated ad-molecule. According to previous reports, the contributions from magnetic dipolar and direct exchange couplings can be negligible due to the large inter-molecular distance (1.56 nm) in the island [28,41]. Therefore, the observed broadening in the Kondo resonance width can be ascribed to the antiferromagnetic RKKY interaction in the many-body Kondo system [41-43].

Figure 4(c) displays the site-dependent *dI/dV* spectra along the red arrow shown in the inset, in which the Kondo resonance exhibits good consistency and spatial recognition in a well-ordered molecular island. Given these observations, each CoPc in the molecular island can be exploited as an information storage bit, and the corresponding memory capacity of the CoPc island is extraordinarily high, on the order of ~5 TB/cm$^2$, which may find far-reaching technological applications.

In summary, we discovered pronounced Kondo resonance of CoPc adsorbed on the semimetallic Sb(111) substrate. Our STM/STS results showed that the spin of a CoPc molecule can endure the adsorption on the substrate, inducing complex spin polarization inside the molecule where the pyrrole ring becomes spin active as well. Our DFT calculations further revealed that the optimal density of states of Sb(111) around $E_F$ plays a crucial role in the preservation of the intrinsic magnetic moment of CoPc while still ensuring sufficiently effective multiple spin-flip scatterings of the conduction electrons of the semimetallic substrate. Moreover, our STS measurements revealed that each of the CoPc molecules in a molecular island exhibits good consistency and spatial recognition of the Kondo resonance, rendering the systems immense potential for applications in ultrahigh-density memory devices. The delicate interfacial physics elucidated in the present study is expected to be also broadly applicable to other related systems combining different magnetic molecules and substrates, including magnetic Weyl semimetallic substrates, serving as ideal platforms for exploring exotic quantum many-body correlations and spintronic phenomena [44,45].

We acknowledge funding from National Key Research and Development Program of China (Grant No. 2017YFA0303500), National Natural Science Foundation of China (Grant Nos. 11634011, 11974323, 12074099), Strategic Priority Research Program of Chinese Academy of Sciences (Grant No. XDB30000000), the Plan for Leading Talent of Fundamental Research of the Central China in 2020, and Intelligence Introduction Plan of Henan Province in 2021 (CXJD2021008), Anhui Initiative in Quantum Information Technologies (Grant No. AHY170000), Natural Science Foundation for Young Scientists of Henan Province (No. 202300410060), and Key Research Project of Henan Provincial Higher Education (No. 20A140005).

[*]These authors contributed equally to this work

[†]zhangzy@ustc.edu.cn

[‡]wfzhang@henu.edu.cn

[1] S. Sanvito, Chem. Soc. Rev. **40**, 3336 (2011).

[2] L. Bogani, and W. Wernsdorfer, Nat. Mater. **7**, 179 (2008).

[3] E. Moreno-Pineda, and W. Wernsdorfer, Nat. Rev. Phys. (2021). https://doi.org/10.1038/s42254-021-00340-3

[4] A. R. Rocha *et al.*, Nat. Mater. **4**, 335 (2005).

[5] G. Czap *et al.*, Science **364**, 670 (2019).

[6] L. Gu, and R. Wu, Phys. Rev. Lett. **125**, 117203 (2020).

[7] A. Zhao *et al.*, Science **309**, 1542 (2005).

[8] U. G. E. Perera *et al.*, Phys. Rev. Lett. **105**, 106601 (2010).

[9] J. Kügel *et al.*, Phys. Rev. Lett. **121**, 226402 (2018).

[10] S. Javaid *et al.*, Phys. Rev. Lett. **105**, 077201 (2010).

[11] L. Farinacci *et al.*, Phys. Rev. Lett. **125**, 256805 (2020).

[12] C. Rubio-Verdú *et al.*, Phys. Rev. Lett. **126**, 017001 (2021).

[13] X. Li *et al.*, Nat. Commun. **11**, 2566 (2020).

[14] K. V. Raman *et al.*, Nature, **493**, 509 (2013)

[15] M. Mannini *et al.*, Nature, **468**, 417 (2010)

[16] E. Minamitani *et al.*, Phys. Rev. Lett. **109**, 086602 (2012).

[17] J. Brede *et al.*, Phys. Rev. Lett. **105**, 047204 (2010).

[18] X. Chen *et al.*, Phys. Rev. Lett. **101**, 197208 (2008).

[19] L. Gao *et al.*, Phys. Rev. Lett. **99**, 106402 (2007).

[20] E. Minamitani *et al.*, Phys. Rev. B **92**, 075144 (2015).

[21] A. Mugarza *et al.*, Nat. Commun. **2**, 490 (2011).

[22] J. D. Baran *et al.*, Phys. Rev. B **81**, 075413 (2010).

[23] B. W. Heinrich *et al.*, J. Phys. Chem. Lett. **1**, 1517 (2010).

[24] S. Stepanow *et al.*, Phys. Rev. B **83**, 220401 (2011).

[25] F. Schulz *et al.*, Nature Physics **11**, 229 (2015).

[26] X. Chen, and M. Alouani, Phys. Rev. B **82**, 094443 (2010).

[27] See Supplemental Material at http://link.aps.org/supplemental for a description of the experimental details and calculated methods, as well as additional experimental data and discussions.

[28] L. She *et al.*, J. Chem. Phys. **136**, 144707 (2010).

[29] A. Zhao *et al.*, J. Chem. Phys. **128**, 234705 (2008).

[30] U. Fano, Phys. Rev. **124**, 1866 (1961).

[31] V. Madhavan *et al.*, Science **280**, 567 (1998).

[32] K. Nagaoka *et al.*, Phys. Rev. Lett. **88**, 077205 (2002).

[33] A. C. Hewson, *The Kondo problem to heavy fermions* (Cambridge University Press, Cambridge, UK, 1993).

[34] Y. Yu *et al.*, ACS. Nano. **8**, 11576 (2014).

[35] Y. Wang *et al.*, J. Chem. Phys. **141**, 084713 (2014).

[36] A. F. Otte *et al.*, Phys. Rev. Lett. **103**, 107203 (2009).

[37] A. F. Otte *et al.*, Nat. Phys. **4**, 847 (2008).

[38] Z. Hu *et al.*, J. Phys. Chem. C **112**, 13650 (2008).

[39] A. Mugarza *et al.*, Phys. Rev. B **85**, 155437 (2012).

[40] R. Bader, *Atoms in molecules: a quantum theory*, Oxford University Press, USA, 1994.

[41] N. Tsukahara *et al.*, Phys. Rev. Lett. **106**, 187201 (2011).

[42] J. Figgins, and D. K. Morr, Phys. Rev. Lett. **104**, 187202 (2010).

[43] P. Wahl *et al.*, Phys. Rev. Lett. **98**, 056601 (2007).

[44] D. F. Liu *et al.*, Science **365**, 1282 (2019).

[45] N. Morali *et al.*, Science **365**, 1286 (2019).

**Figures and captions**

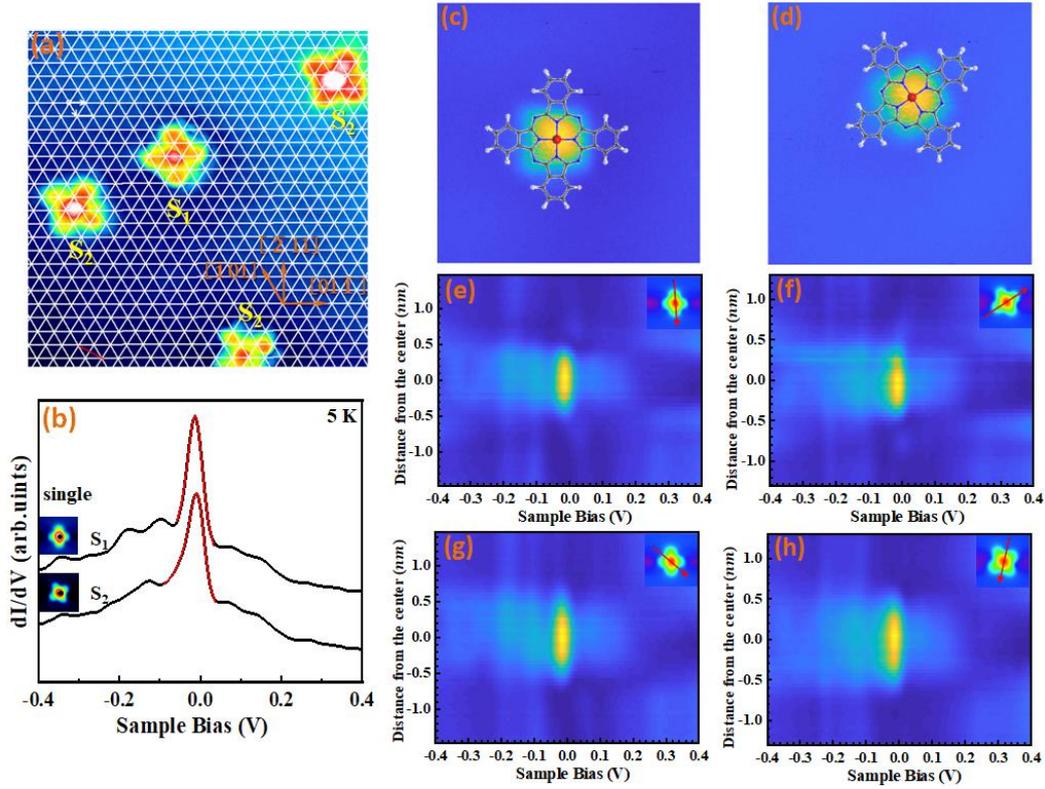

FIG. 1. (a) High resolution STM image (9.1 × 9.1 nm$^2$; $U = 0.4$ V; $I = 300$ pA) with the isolated CoPc molecules and substrate lattice resolved simultaneously. The white grid represents the substrate lattice [see Fig. S1(a) in SM [27]], showing two configurations ($S_1$ and $S_2$), where the molecules are adsorbed at the bridge site of the substrate. The crystallographic directions and unit cell of the Sb(111) substrate are denoted by the orange and white arrows, respectively. (b) dI/dV spectra for the two configurations, acquired on the Co center of the isolated molecules on Sb(111) at 5 K. (c) and (d) dI/dV mapping for both configurations, taken at -10 and -3 mV respectively (4.0 × 4.0 nm$^2$; $I = 100$ pA). (e)-(h) Spatially dependent dI/dV spectra of the isolated CoPc molecules acquired along the lines marked by the red arrows in the inserts.

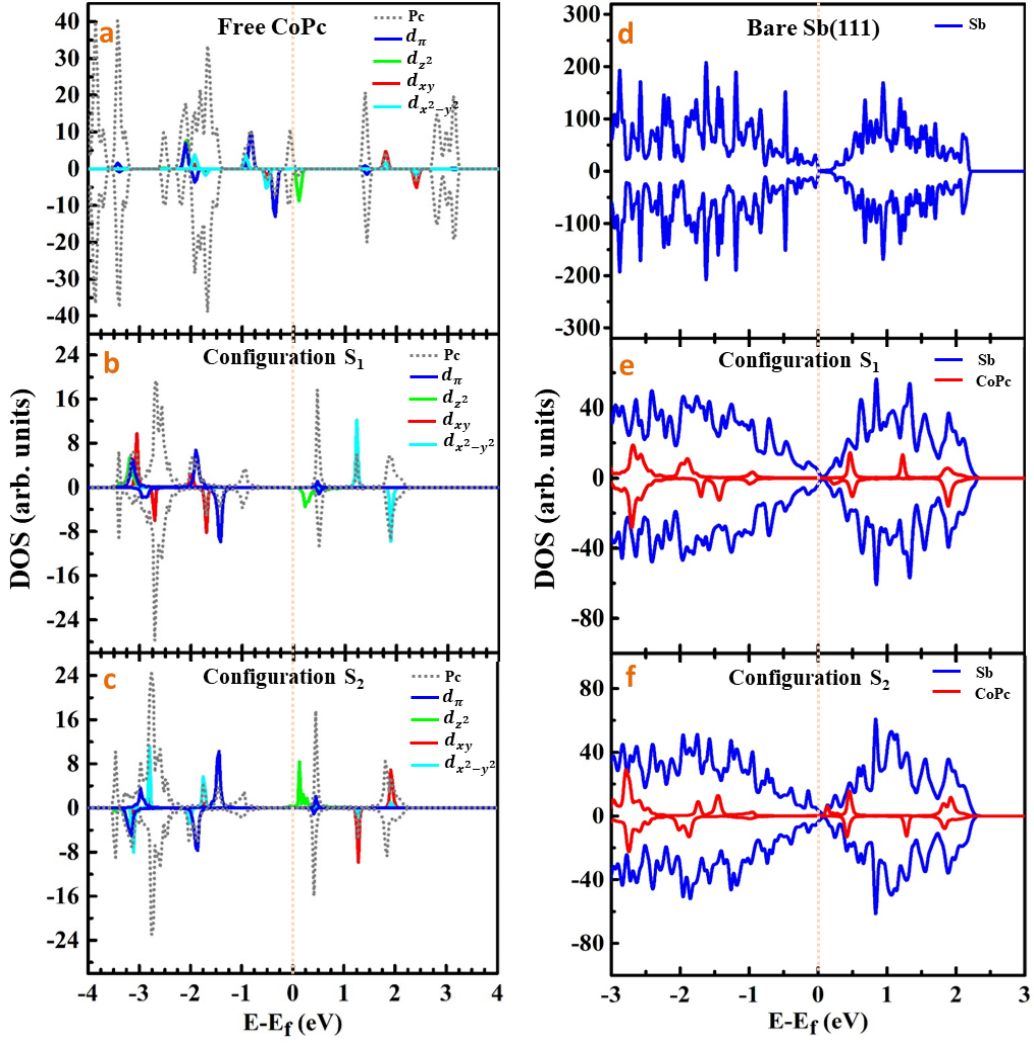

FIG. 2. (a)-(c) Spin-polarized partial density of states of Co and Pc in the free CoPc, $S_1$, and $S_2$ configurations, respectively. The gray, blue, green, red, and cyan curves represent the DOSs of Pc, $d_\pi$, $d_{z^2}$, $d_{xy}$, and $d_{x^2-y^2}$, respectively. (d)-(f) Partial density of states of bare Sb(111) and CoPc/Sb(111) composites. The blue and red curves represent the DOSs of Sb and CoPc, respectively.

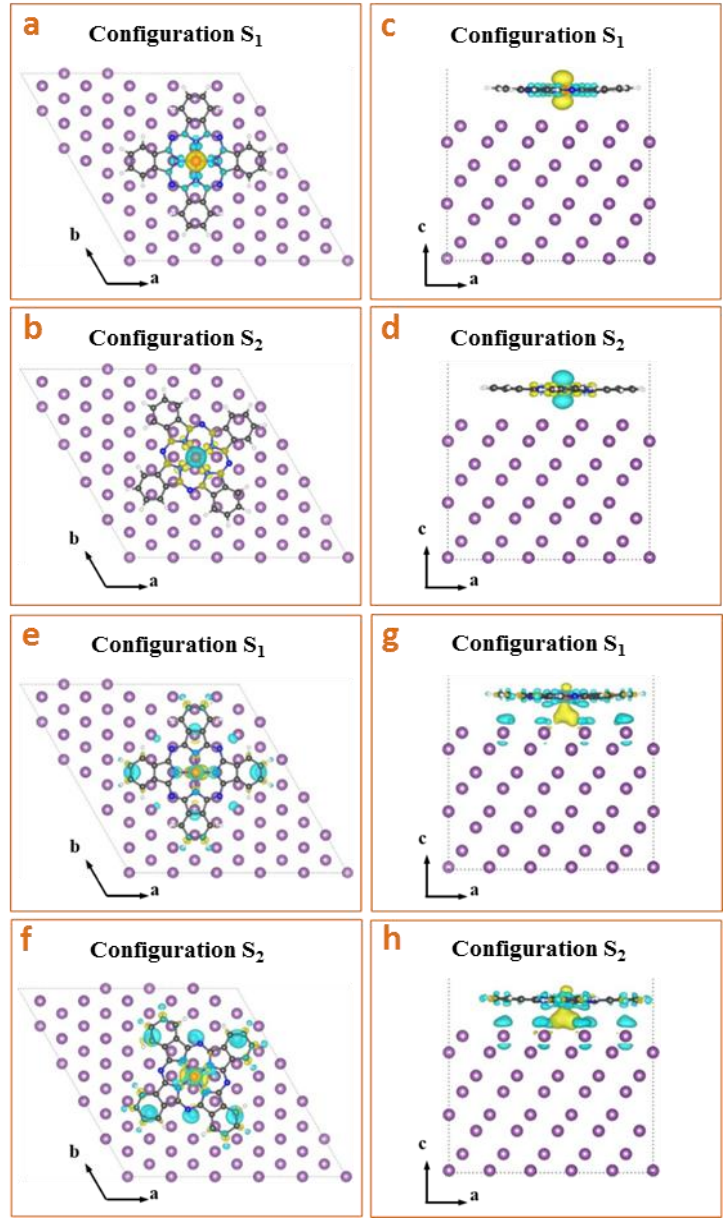

FIG. 3. (a)-(d) Top and side views of the spin distributions for configurations $S_1$ and $S_2$. The spin-up and spin-down contributions are denoted in yellow and cyan, respectively. The isovalue for the spin density is 0.0006 e/Å$^3$. (e)-(h) Top and side views of differential electronic density maps for configurations $S_1$ and $S_2$. The yellow and cyan colors represent increase and decrease in the electron density, respectively. The isovalue of the electron density is 0.0004 e/Å$^3$.

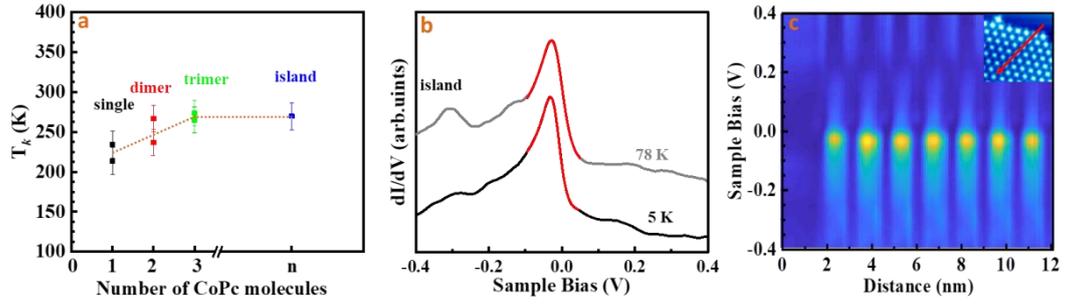

FIG. 4. (a) Kondo temperature as a function of the number of the CoPc assembly size. (b) $dI/dV$ spectra taken on the Co center of the CoPc molecules in a molecular island at 5 and 78 K. The red curves depict the Fano fittings to the Kondo resonances. (c) Site-dependent $dI/dV$ spectra of the molecular island acquired along the line marked by the red arrow in the upper-right inset.

# Supplementary Materials for
# Magnetic moment preservation and emergent Kondo resonance of Co-phthalocyanine on semimetallic Sb(111)

by She *et al.*

## List of contents

**Supplementary Section I: Experimental details**

The Sb(111) substrate were cleaned by repeated Ar$^+$ sputtering and annealing at about 620 K. Before evaporating from a quartz crucible onto the Sb(111) substrate, CoPc powder (purity of 95%, Aldrich) was degassed slightly above the required deposition temperature (~600 K) for several hours to remove impurities. After that, the CoPc were evaporated onto the Sb(111) substrate with a rate of ~0.1 ML/min at room temperature. All the *dI/dV* spectra and *dI/dV* maps were acquired by lock-in technique with sinusoidal modulations (877 Hz) of 10 and 2 mV, respectively.

**Supplementary Section II: Computational details**

Electronic structure calculations were performed within density functional theory as implemented in the Vienna ab initio simulation package, employing projected augmented wave potentials to describe the atomic core electrons and a plane wave basis set with a kinetic energy cutoff of 400 eV to expand the Kohn−Sham electronic states [1,2]. For the exchange and correlation functional, the generalized gradient approximation (GGA) in the Perdew–Burke–Ernzerhof (PBE) format was used [3]. To properly describe the correlation effects of the 3d electrons, the GGA + *U* ($U_{eff}$ = 2.0 eV) method was adopted [4,5]. A periodic slab model consisting of 257 atoms was employed in the calculations, which was constructed by four Sb bilayers with a (5×5) Sb(111) surface primitive cell, and one CoPc molecule was adsorbed on the bridge site. A vacuum region of 20 Å in the *z* direction was applied to avoid interaction between image slabs. The Brillouine zone was sampled in (3×3×1) and (7×7×1) *k*-point meshes in structure relaxations and DOS calculations, respectively. To better account for the intermolecular interactions between the Sb substrate and CoPc molecule, the DFT-D3 method with Becke-Jonson damping was adopted [6]. During structural relaxation, all the structures were optimized until the force on each atom was less than 0.02 eV Å$^{-1}$. The spin-polarization effect was included in all the calculations.

**Supplementary Section III: CoPc adsorption geometry on Sb(111)**

Supplementary Figure S1(a) is a STM image (5 K) with CoPc molecules and the substrate lattice resolved simultaneously, showing the isolated CoPc molecule located at the bridge site of the Sb(111) substrate. In comparison, the CoPc molecule is rotating and easily dragged by the STM tip with a high tip-height at 78 K (see Supplementary Fig. S1(b)). Due to the weak van der Waals interaction, the isolated molecule is difficult to stabilize on the Sb(111) surface at a temperature of tens of Kelvin.

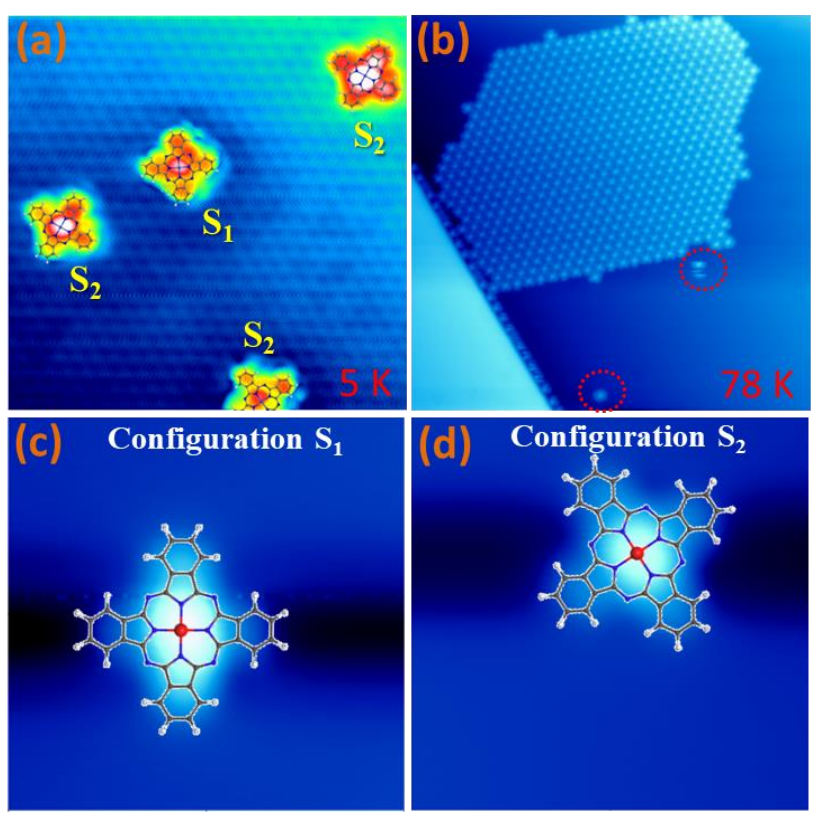

Fig. S1. (**a**) High resolution STM image of isolated CoPc on Sb(111) was acquired at 5 K (9.1 × 9.1 nm$^2$; $U$ = 0.4 V; $I$ = 300 pA). (**b**) Largely scaled image of CoPc molecular island was obtained at 78 K (52.1 × 52.1 nm$^2$; $U$ = 2.8 V; $I$ = 20 pA). As labeled by red broken circles, isolated CoPc is rotating on Sb(111) surface and easily dragged by the STM tip. (**c-d**) STM images of configurations S$_1$ and S$_2$ were simultaneously acquired with $dI/dV$ mapping, respectively, indicating that the sharp tip has a very well spatial resolution during STM measurements.

**Supplementary Section IV: Spatial dependent half-width of Kondo resonance**

For configurations $S_1$ and $S_2$, a series of *dI/dV* spectra along the red arrows in the inset were fitting with the Fano function. The extracted half-widths of Kondo resonances at Co and pyrrole ring are identical, indicating the same Kondo temperature, as shown in Supplementary Fig. S2(a),(b).

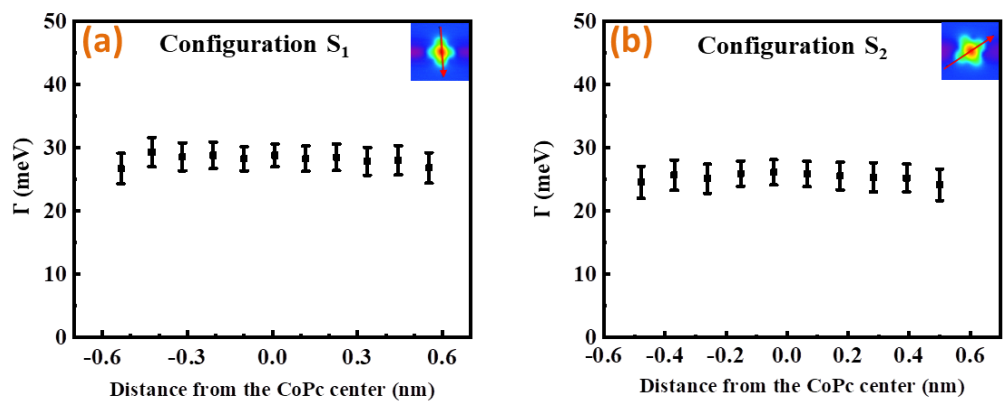

Fig. S2. Half-width of Kondo resonance extracted from the experimentally measured *dI/dV* spectra along the red arrows in the upper-right inset, (**a**) configuration $S_1$, and (**b**) configuration $S_2$.

**Supplementary Section V: The apparent height of CoPc on Sb(111)**

In order to reduce the impact of the electronic density of states of Co ion on the apparent distance of CoPc on Sb(111), the topographic images of both configurations obtained at small positive biases, as shown in Supplementary Fig. S3(a) and S3(b). The Co-Sb distance is roughly obtained by the apparent height measured from the topographic images. Supplementary Fig. S3(c) and S3(d) show the height profiles along red line through the center of CoPc molecules, indicating that the apparent height of the CoPc is about 3.2 Å, agreeing well with DFT calculations.

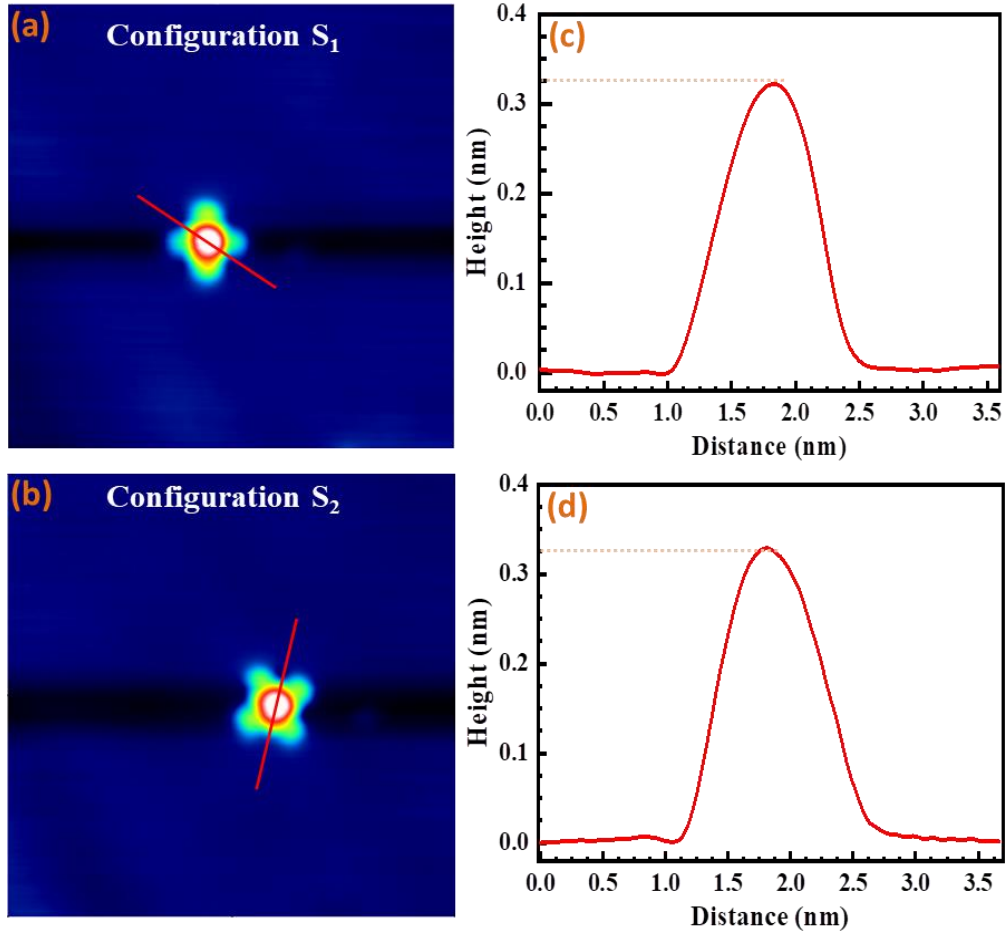

Fig. S3. (**a**),(**b**) STM images of the isolated CoPc (at 5 K) were obtained at bias voltage 0.1 V ((10 × 10 nm$^2$, I = 30 pA ). (**c**), (**d**) Height profile taken along the red line in (a) and (b), showing that the apparent high of the CoPc on Sb(111) is about 3.2 Å.

## Supplementary Section VI: Magnetic moment of CoPc/Sb(111) as a function of Co-Sb distance

To investigate the influence of Co-Sb distance on the magnetic moment of CoPc-Sb(111) system, the CoPc molecule has been pulled close to the Sb(111) surface artificially from the equilibrium distance. As shown in Supplementary Fig. S4, the magnetic moment of CoPc disappears rapidly as the Co-Sb distance decreases, which implies that a proper Co-Sb distance is crucial to preserve the molecular spin.

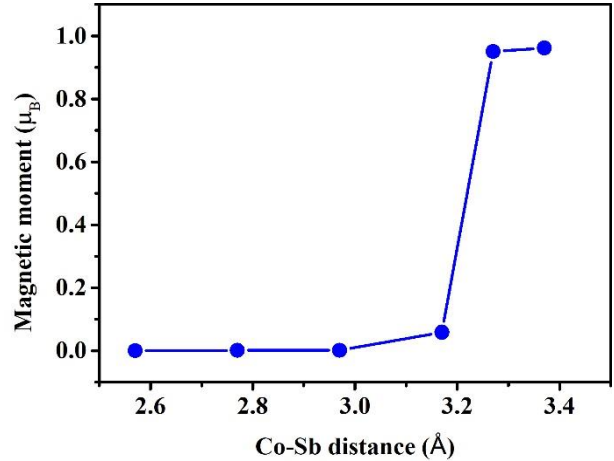

Fig. S4. Magnetic moment of CoPc/Sb(111) at configuration $S_1$, as a function of Co-Sb distance. The equilibrium distance between Co and Sb(111) surface is 3.37 Å.

**Supplementary Section VII:** *dI/dV* spectra of the dimer and trimer

To reveal the evolution of Kondo resonance from a single CoPc molecule to the two-dimensional island, we measured *dI/dV* spectra of the CoPc dimer and trimer (see Supplementary Fig. S5(a) and S5(b)). Compared with that of the isolated CoPc, the fitting widths of the Kondo resonance in the dimer and trimer increase with the number of the molecules. Our fitting results indicate that the Kondo temperature of the trimer is very close to that of CoPc in molecular island.

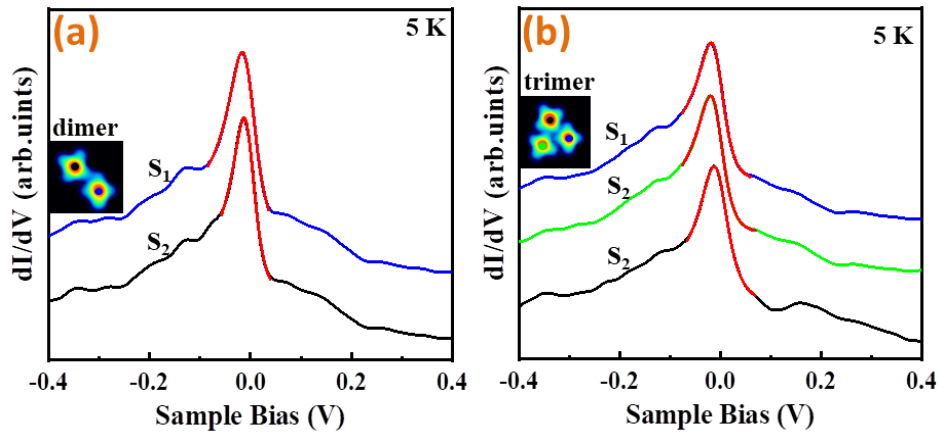

Figure S5. (**a**), (**b**) *dI/dV* spectra for the dimer and trimer, acquired on the Co center of the CoPc molecules on Sb(111) at 5 K. The color of *dI/dV* curves corresponds to that of dots (measurement positions) in the insets. The red lines are fits of the *dI/dV* curves to the Fano function.

**Supplementary References**

[1] G. Kresse, and J. Hafner, Ab initio molecular dynamics for liquid metals, Phys. Rev. B **47**, 558(R) (1993).

[2] G. Kresse, and J. Furthmüller, Efficient iterative schemes for ab initio total-energy calculations using a plane-wave basis set, Phys. Rev. B **54**, 11169 (1996).

[3] J. P. Perdew, K. Burke, and M. Ernzerhof, Generalized gradient approximation made simple, Phys. Rev. Lett. **77**, 3865 (1996).

[4] S. L. Dudarev, G. A. Botton, S. Y. Savrasov, C. J. Humphreys, and A. P. Sutton, Electron-energy-loss spectra and the structural stability of nickel oxide: An LSDA+U study, Phys. Rev. B **57**, 1505 (1998).

[5] Y. Wang, X. Li, and J. Yang, Electronic and magnetic properties of CoPc and FePc molecules on graphene: the substrate, defect, and hydrogen adsorption effects, Phys. Chem. Chem. Phys. **21**, 5424 (2019).

[6] S. Grimme, S. Ehrlich, and L. Goerigk, Effect of the damping function in dispersion corrected density functional theory, J. Comp. Chem. **32**, 1456 (2011).